\documentclass[useAMS,usenatbib]{mn2e}
\usepackage{graphicx, color}
\usepackage{aas_macros}

\usepackage{braket}
\usepackage{bm}
\usepackage{threeparttable}
\usepackage{amssymb, amsmath}
\usepackage{multirow}

\newcommand{\Msun}{\rm M_{\odot}}

\newcommand{\Ha}{H{$\rm \alpha$}~}
\newcommand{\OIII}{O{\sc iii}}

\title[Deep learning for intensity mapping]
{Deep learning for intensity mapping observations: Component extraction}

\author[K. Moriwaki et al.]
{Kana Moriwaki$^{1}$\thanks{E-mail: kana.moriwaki@phys.s.u-tokyo.ac.jp},
Nina Filippova$^{1,2}$, 
Masato Shirasaki$^{3}$,
Naoki Yoshida$^{1, 5, 6, 7}$
\\
$^{1}$Department of Physics, The University of Tokyo, 7-3-1 Hongo, Bunkyo, Tokyo 113-0033, Japan \\
$^{2}$Department of Physics, Princeton University, Princeton, NJ 08544, USA\\
$^{3}$National Astronomical Observatory of Japan (NAOJ), Mitaka, Tokyo 181-8588, Japan\\
$^{4}$Kavli Institute for the Physics and Mathematics of the Universe (WPI), 
UT Institutes for Advanced Study, \\
The University of Tokyo, 5-1-5 Kashiwanoha, Kashiwa, Chiba 277-8583, Japan \\
$^{5}$Research Center for the Early Universe, School of Science, The University of Tokyo, 
7-3-1 Hongo, Bunkyo, Tokyo 113-0033, Japan \\
$^{6}$Institute for Physics of Intelligence, School of Science, The University of Tokyo, 
7-3-1 Hongo, Bunkyo, Tokyo 113-0033, Japan \\
}

\begin{document}

\date{}

\pagerange{\pageref{firstpage}--\pageref{lastpage}}
\pubyear{0000}

\maketitle

\label{firstpage}

\begin{abstract} 
  Line intensity mapping (LIM) is an emerging observational method to study the large-scale structure of
  the Universe and its evolution. 
  LIM does not resolve individual sources but probes the fluctuations of integrated line emissions.
  A serious limitation with LIM is that contributions of different emission lines from sources at different redshifts are all confused at an observed wavelength.
  We propose a deep learning application to solve this problem.
  We use conditional generative adversarial networks to extract designated information from LIM. We consider a simple case with two populations of emission line galaxies; \Ha emitting galaxies at $z = 1.3$ are confused with 
  [\OIII] emitters at $z = 2.0$ in a single 
  observed waveband at 1.5 $\rm \mu m$.
  Our networks trained with $30,000$ mock observation maps are able to extract the total 
  intensity and the
  spatial distribution of \Ha emitting galaxies at $z = 1.3$.
  The intensity peaks are successfully located with 74\% precision.
  The precision increases to 91\% when we combine 5 networks.
  The mean intensity and the power spectrum are reconstructed with an accuracy of $\sim$10\%.
  The extracted galaxy distributions at a wider range of redshift can be used for studies on 
  cosmology and on galaxy formation and evolution.
\end{abstract} 


\section{introduction}

Line intensity mapping (LIM) is a promising observational technique for next-generation
cosmology. LIM probes the large-scale structure of the Universe  
at a wide range of redshift and thus enables us to study cosmology as well as galaxy formation and evolution
\citep[][]{Kovetz17}.
Fluctuations of the integrated intensity of emission lines such as Lyman-$\alpha$, \Ha, [C{\sc ii}],
and CO lines trace the distributions of the underlying galaxies,
while hydrogen 21-cm line is used to study the distribution and ionization state of the intergalactic medium in the early Universe \citep{Pritchard12}. 

A number of LIM observation programmes have been proposed and are planned \citep[see][]{Kovetz17}.
LIM measures the integrated emission from all the sources redshifted to a wavelength bin.
While it provide rich information on the sources and their large-scale distribution in principle, 
the confusion of sources or contamination
from foreground/background emission is an inevitable problem in practice.
\citet{Fonseca17} show that multiple emission lines from galaxies often contribute 
roughly equally to the total intensity at a certain observed wavelength.
There are a few methods to infer the contribution from a designated redshift.
One is to perform cross-correlation analysis with other known tracers
of galaxies or of the matter distribution at the same redshift \citep[e.g.][]{Visbal10}.
More practical methods such as masking brightest pixels allow 
to detect subdominant signals \citep{Gong14, Silva18}.
It is also possible to distinguish signals from different redshifts
using the anisotropic power spectrum shapes \citep[e.g.][]{Cheng16}.
These methods are aimed at estimating the statistical quantities,
but do not generate direct images of the intensity distribution.  
It would be more informative and useful if contaminants are removed 
from an image to show explicitly the intensity distribution at an arbitrary redshift.
Here, we propose to use deep learning to separate/extract information
from intensity maps.

Convolutional neural networks (CNNs) are a popular and promising tool
for image processing including problems related to LIM.
Recent studies propose to use CNNs to analyse hydrogen 21-cm line signals
from the epoch of reionization \citep{Hassan19a, Hassan19b, Gillet19, Zamudio-Fernandez19} 
or to estimate the line luminosity function from a CO intensity map \citep{Pfeffer19}.
\citet{Shirasaki19} use conditional generative adversarial networks 
\citep[cGANs, e.g.,][]{Isola16} to de-noise observed weak-lensing mass maps.
A cGAN consists of a pair of CNNs that learn an image-to-image translation in an adversarial way, and is able to generate fine and complicated images.

In this letter, we apply cGANs to intensity maps
to reconstruct the intensity distribution and basic statistics of galaxy distribution.
We aim at decoding cosmological information
from future intensity map observations
using ground-based and space-borne telescopes.  
We show that our networks, after appropriately trained with a large set of
mock observations, can generate accurately the intensity distribution from a single source population.
Throughout this letter, we adopt $\rm \Lambda$CDM cosmology with
$\Omega_{\rm M} = 0.316, \Omega_{\rm \Lambda} = 0.684, h = 0.673$ \citep{Planck18}.

\section{methods}

We consider the line intensity observed at wavelength of $1.5~ \rm \mu m$. 
Planned or proposed near-infrared LIM projects include
the Spectrophotometer for the History of the Universe, Epoch of Reionization, and Ice Explorer \citep[SPHEREx,][]{Dore16} 
and the Cosmic Dawn Intensity Mapper \citep[CDIM,][]{Cooray19}.
Emission lines from galaxies at $z \sim 0-5$ are considered to be the dominant
sources in this spectral regime.
As a simple but realistic case, we assume that the observed line intensity
map consists of two most dominant emission lines:
\Ha line from $z = 1.3$ and [\OIII] 5007\AA~line from $z = 2.0$. 
Observational noises and other contaminants such as [O{\sc ii}] 3727\AA~are to be considered in a forthcoming paper (Moriwaki et al. in preparation).

\subsection{Mock intensity maps}

We generate a number of mock intensity maps for training and testing in the following manner.
First, we populate a cubic volume of $280h^{-1}~\rm Mpc$ with
dark matter haloes using the publicly available {\sc pinocchio} code \citep{Monaco13}. 
We set the minimum halo mass of the catalog to be $3\times 10^{10}h^{-1}~\Msun$. We have tested and confirmed that the map properties such as the total line intensity are not significantly
affected by this choice of the minimum halo mass.

We derive the halo mass-luminosity relation using the outputs of the cosmological hydrodynamics simulation Illustris-TNG \citep{Nelson19}.
We use the TNG300-1 dataset which has a simulated volume of $V_{\rm box} = (302.6~\rm  cMpc)^3$.
We compute the line luminosity from a simulated galaxy as
\begin{eqnarray}
	L_{\rm line} = 10^{-A_{\rm line}/2.5}\, C_{\rm line}(Z)\, {\rm SFR},
\end{eqnarray}
where $A_{\rm line}$ accounts for attenuation by dust.
We adopt $A_{\rm H\alpha} = 1.0$ mag and $A_{\rm [O{\sc iii}]} = 1.35$ mag. We use the photoionization simulation code {\sc cloudy} \citep{Ferland17} to compute the coefficient $C_{\rm line}(Z)$ 
as a function of the mean metallicity of the galaxy.
The {\sc cloudy} computation is done in the same manner as in \citet{Moriwaki18}
except that we adopt typical values 
of the electron density $n = 100~\rm cm^{-3}$ and the ionization parameter $U = 0.01$.
We compute the luminosity of a simulated halo by summing up the luminosities of its member galaxies.
The halo mass-luminosity relation, i.e., the mean $\overline{L}_i$ and the variance $\sigma_i$ within each halo mass bin, is then obtained
from the Illustris output.
To generate an emissivity field from a {\sc pinocchio} halo catalogue, we assume a Gaussian distribution with a mean $\overline{L}_i$ and a variance $\sigma_i$ and assign luminosities to the haloes in $i$-th  mass bin. 

We perform the above procedure for \Ha and [\OIII] indenpendently at the respective redshift. 
We generate two-dimensional intensity maps by projecting the three-dimensional emissivity fields along one direction.
The total area subtended by a map is $(3.4 ~\rm deg)^2$, and we assume a spectral resolution $R = 40$ that corresponds to the expected resolution of SPHEREx.
We find that the relative contribution from the [\OIII] emission (at $z=2.0$) is $\sim 60\%$ of the \Ha map (at $=1.3$), which is consistent with other theoretical studies \citep{Fonseca17, Silva18}.
To generate a large number of training data set, we use 300 different halo catalogues.

For each realization, 
100 maps with an area of $(1.7 ~\rm deg)^2$ are generated by projecting along random direction. We obtain 30,000 training data in total.
In this way, we obtain training maps with various mean intensities.
Each map has $256\times 256$ pixels, corresponding to a pixel size of $(0.4 ~\rm arcmin)^2$. 
For the test data set, we produce another 1,000 halo catalogs and generate 1000 independent maps.
We smooth the training and test maps with a Gaussian beam with $\sigma = 1.2~\rm arcmin$ before giving them to the networks.\footnote{We have performed the same analysis in the present paper to the "raw" data without smoothing. We have found the performance of the cGANs is somewhat degraded when the networks are trained with the unsmoothed data populated with a number of discrete sources.
}

\begin{figure*}
\begin{center}
\includegraphics[width=13.7cm]{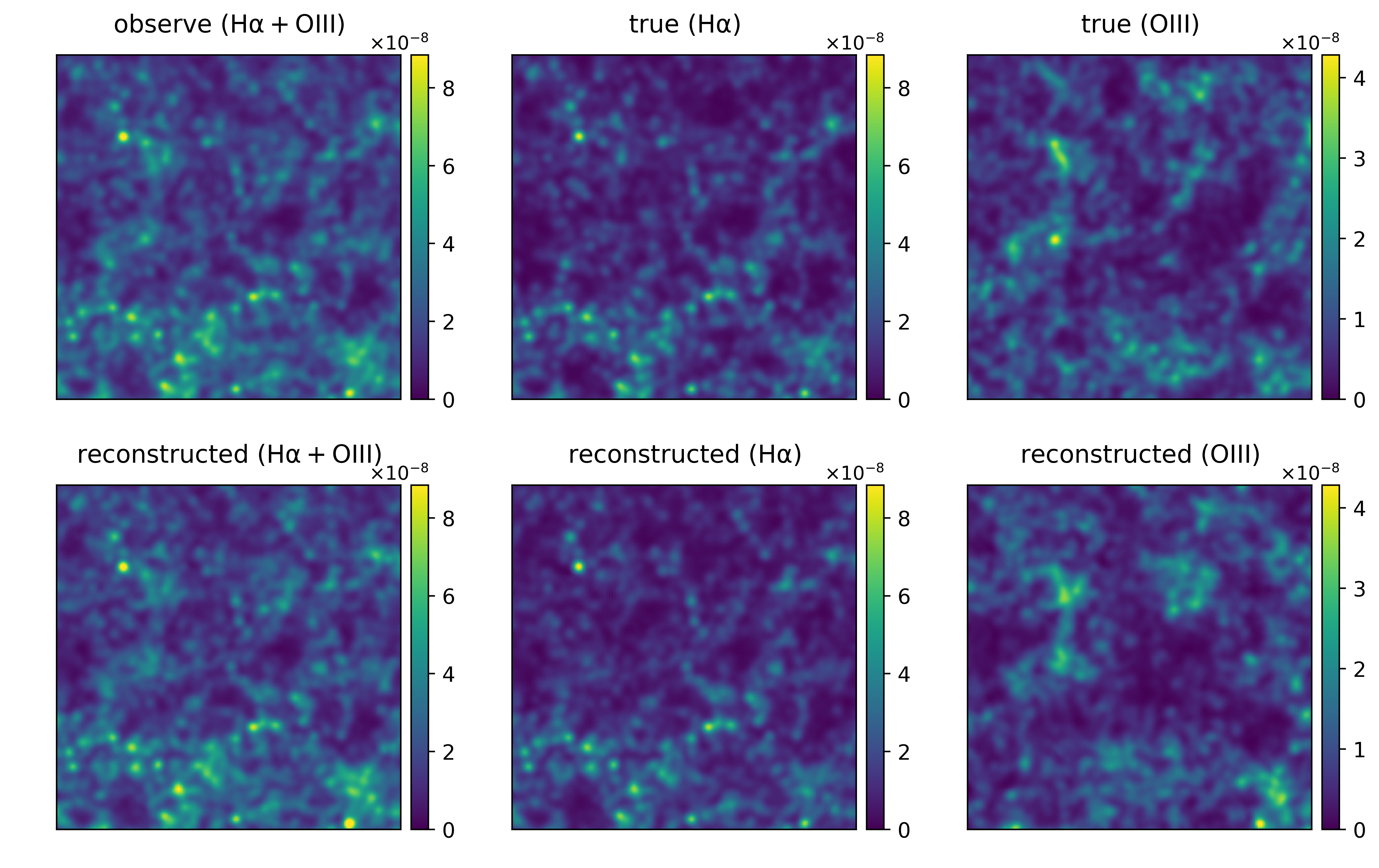}
\caption{
An observed map (top-left) is contributed by \Ha (top-center) and [\OIII] (top-right) emission.
The reconstructed \Ha and [\OIII] maps and the sum of them are shown in bottom.
The shown area is $1.7 ~\rm deg$ on a side, and the intensities are in units of $\rm erg~s^{-1}~cm^{-2}~sr^{-1}$. Note the relative difference in intensity for \Ha and [\OIII] (colour bars). 
Our network reconstructs even the fainter [\OIII] component.
}
\label{fig:map}
\end{center}
\end{figure*}

\subsection{Convolutional networks}

We construct cGANs based on {\sc pix2pix} by \citet{Isola16}.\footnote
{
https://github.com/yenchenlin/pix2pix-tensorflow
}
We train the networks so that they reconstruct both \Ha and [\OIII] images from an observed image.
This kinds of one-to-many image translation networks are studied by, for instance, \citet{Lee18} for separating transparent and reflection scenes.

We have two pairs of adversarial convolutional networks called generator and discriminator.
They are denoted by $(G_1, D_1)$ for \Ha map and by $(G_2, D_2)$ for [\OIII] map.
The generators try to reconstruct \Ha and [\OIII] maps from an observed map $X_{\rm obs}$,
whereas the discriminators try to distinguish the true maps $X_{\rm true,i}$
and the reconstructed maps $G_i(X_{\rm obs})$.
In other words, for an input $(X_{\rm obs}, X)$ with $X$ being either $X_{\rm true, i}$ or $G_i(X_{\rm obs})$, 
the discriminator returns a probability that $X$ is $X_{\rm true, i}$.
Here, $X_{{\rm true}, i} (i = 1,2)$ denote the true maps of \Ha and [\OIII], respectively. 
The generator consists of 8 convolution layers
followed by 8 de-convolution layers
and the discriminator consists of 4 convolution layers.
Two generators $G_1$ and $G_2$ share the first 8 convolution layers.
The kernel size of the convolutions is $5 \times 5$.
In each layer, batch normalization,\footnote{
During test phase, we set {\tt is\_training} = False in batch normalization to use fixed normalization parameters.
} dropout, and skip connection are also performed \citep[see][for more details]{Isola16}.

During the training phase, the performance of the generators and the discriminators are evaluated
by a linear combination of the cross-entropy losses and the mean L1 norms:
\begin{eqnarray}
	\mathcal{L} &=& \sum_{i = 1,2} [ \mathcal{L}_{\rm cGAN}(G_i,D_i) 
	+ \lambda_i\, \mathcal{L}_{\rm L1}(G_i)] \nonumber \\
	&& +\, \lambda_{\rm tot}\, \mathcal{L}_{\rm L1, tot}(G_1,G_2), \label{eq:loss}
\end{eqnarray}
where
\begin{eqnarray}
	\mathcal{L}_{\rm cGAN}(G_i,D_i) &=& \log D_i(X_{\rm obs}, X_{{\rm true},i}) \nonumber \\
	&& +\, \log [1 - D_i(X_{\rm obs}, G_i( X_{\rm obs}))],	
\end{eqnarray}
\begin{eqnarray}
	\mathcal{L}_{\rm L1}(G_i) = \frac{1}{N_{\rm pix}}\sum_{\rm map} | X_{{\rm true},i} - G_i(X_{\rm obs}) |,
\end{eqnarray}
\begin{eqnarray}
	\mathcal{L}_{\rm L1, tot}(G_1, G_2) = \frac{1}{N_{\rm pix}}\sum_{\rm map} | X_{\rm obs} - G_1(X_{\rm obs}) - G_2(X_{\rm obs})|,
\end{eqnarray}
where $N_{\rm pix} = (256)^2$.
In each training set, the generators (discriminators) are updated
to decrease (increase) the loss function $\mathcal{L}$
averaged over a mini batch.\footnote{
Mini batch is a randomly selected set of training data, $\{X_{{\rm obs},i}, X_{{\rm true},i}\}_{i=0}^{n_{\rm b}}$, where $n_{\rm b}$ is batch size.
In training phase, the networks pass through all the training data without duplication.
When we set the number of epochs $n_{\rm e} > 1$, this passing through is repeated for $n_{\rm e}$ times.
For $n_{\rm d}$ training data,
updates are performed for $n_{\rm d}n_{\rm e}/n_{\rm b}$ times in total.
}
We adopt $\lambda_1 = \lambda_2 = \lambda_{\rm tot} = 100$ and a batch size of 4.
The networks are trained for 8 epochs.
We use the Adam optimizer \citep{Kingma14} 
with learning rate 0.0002, and decay rate parameters $\beta_1 = 0.5$ and $\beta_2 = 0.999$ for updating the parameters.

\section{results}

\subsection{Intensity reconstruction}

We study the performance of our networks with 1000 test data.  
Fig. \ref{fig:map} shows an example of true and reconstructed maps.
In our fiducial case of $\lambda_{\rm obs} = 1.5 \mu$m,
the contribution from \Ha is larger than [\OIII], 
and then outstanding structures
in the observed map mostly originate from the \Ha emission at $z=1.3$.
It is thus remarkable that not only the \Ha distribution but also the
weaker [\OIII] intensity is reproduced well.

It is important to study whether statistical quantities are also reproduced accurately.
We first examine the peaks in our intensity maps.
We select as "peaks" local maxima with heights greater than $3\sigma$.
We find 24089 (18859) and 24800 (17631) peaks in the true and
the reconstructed \Ha ([\OIII]) maps over our 1000 test data sets.
Among them, 18262 (5095) peaks are matched correctly.
This means that 76\% (27\%) of the true peaks are 
reproduced, and 74\% (29\%) of the reconstructed peaks are true.

If our purpose is to study individual peaks or other individual structures, we may require much higher accuracy for reconstructed structures.
Previous studies developed cGANs that also learn the reliability of reconstructed maps \citep{Lee18, Kendall17}.
In principle, we can use these methods to 
quantify the reliability of the outputs. 
Another promising idea is to combine multiple networks.
To test this idea, we use 5 networks that have an identical
set of convolutional layers but are trained with different sets of data.
Intensity maps reconstructed by the 5 trained networks 
are similar, but not exactly the same. 
We find that it is generally difficult to reproduce the true intensity
in portions where these networks commonly fail.
For \Ha ([\OIII]) maps, 
the number of peaks detected by all the 5 networks is 14332 (895). 
Among them, 13018 (539) peaks are true, which means a 91\% (60\%) confidence level
for our peak detection.
We note that if we take the average or the median of 
the reconstructed maps 
by multiple networks on a pixel-by-pixel basis, 
dark structures in void regions and small-scale structures 
are smoothed out.

\subsection{Statistical information}

\begin{figure}
\begin{center}
\includegraphics[width=5.7cm]{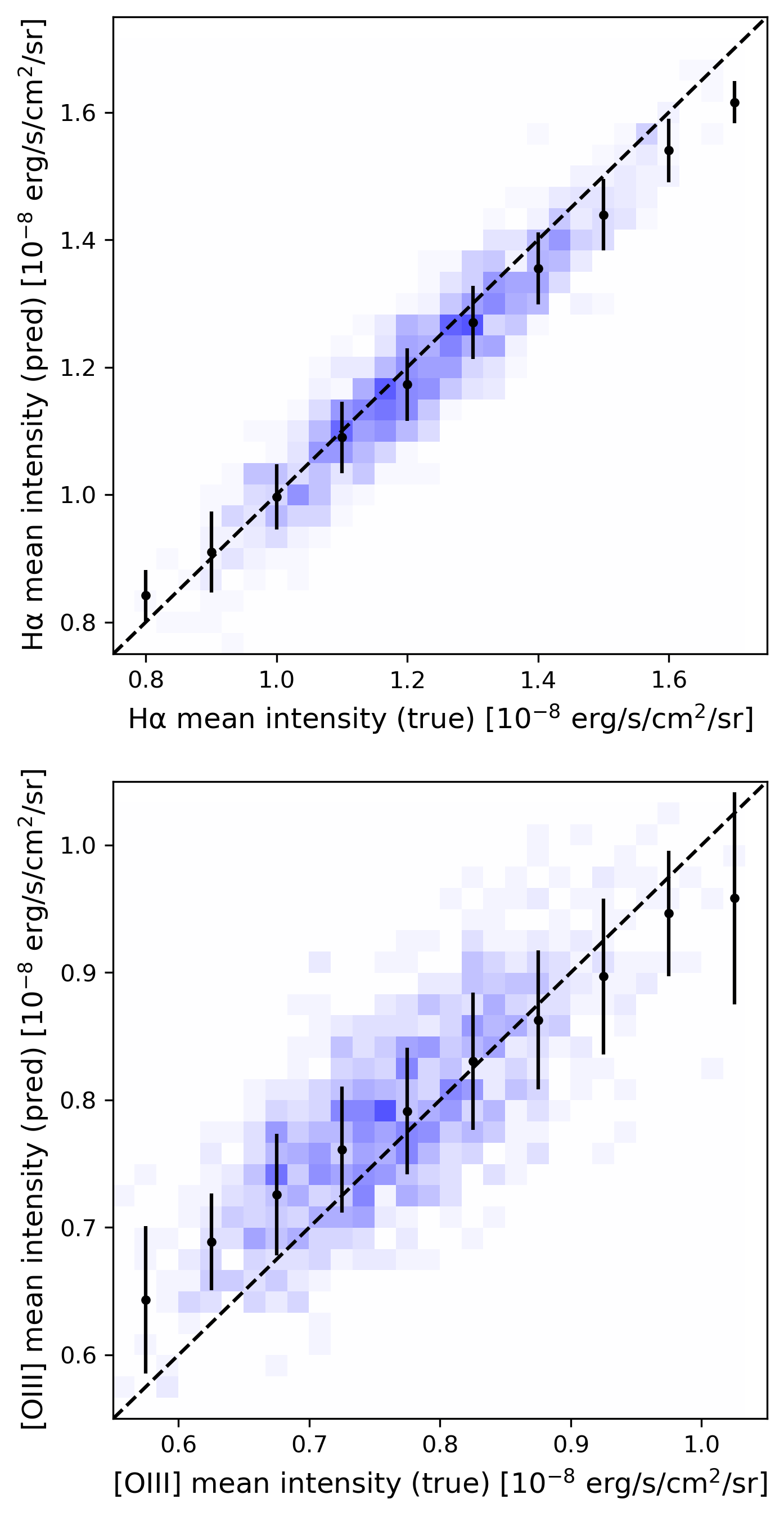}
\caption{
  The mean intensities of the reconstructed maps against the mean of
  the true maps of \Ha (upper) and [\OIII] (bottom)
  for our 1000 test data set.
}
\label{fig:mean_intensity}
\end{center}
\end{figure}

\begin{figure}
\begin{center}
\includegraphics[width=6.7cm]{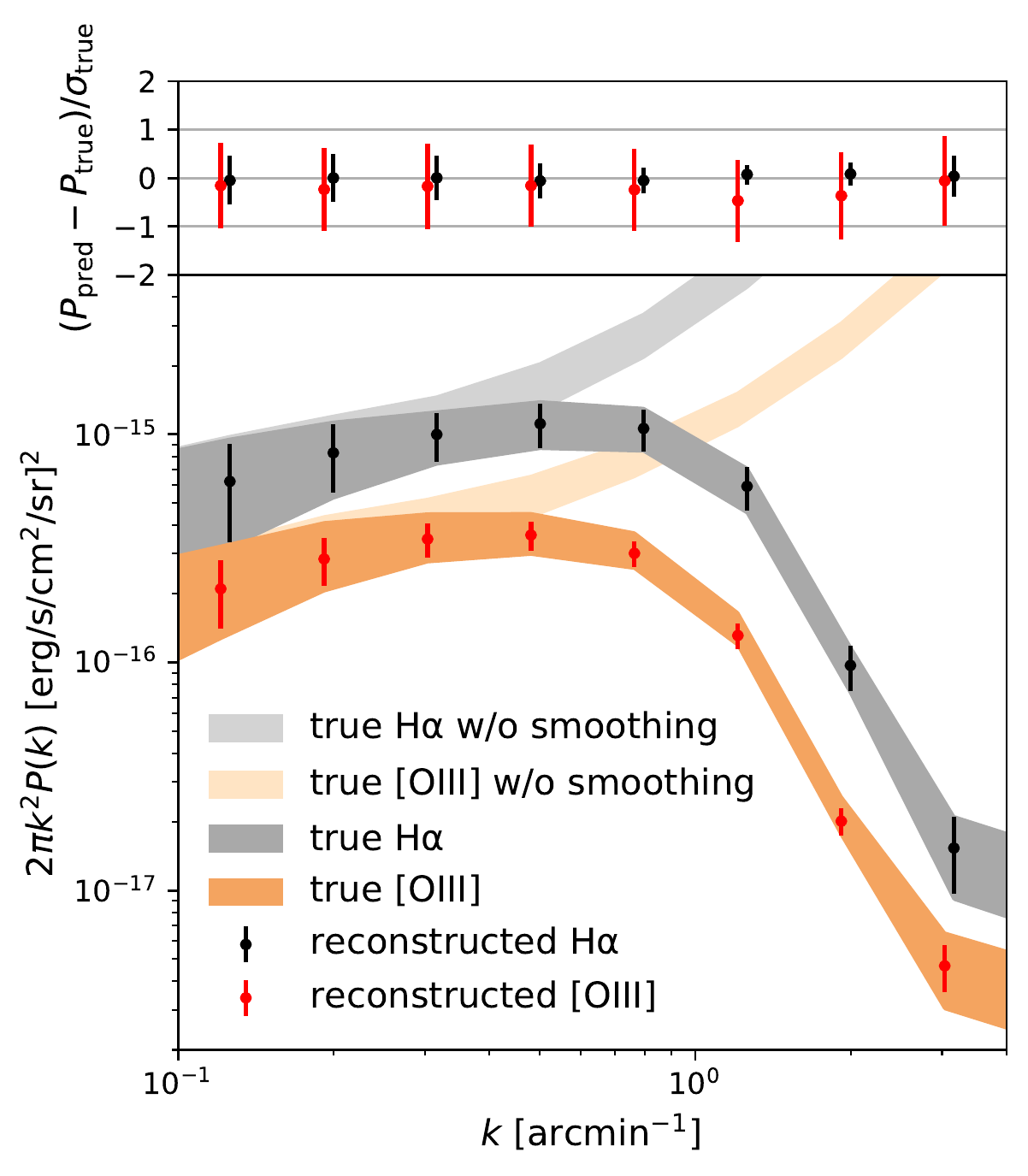}
\caption{
The two-dimensional power spectra of the reconstructed maps.
The error bars and the shaded regions in the bottom panel show the 1 $\sigma$ variance of the power spectrum of the reconstructed and the true maps over 1000 test data, respectively. The light-coloured regions show the 1 $\sigma$ variance of true maps without smoothing.
In the upper panel, we show the difference between the reconstructed and the true power spectra normalized by the variance of the true power spectrum.
}
\label{fig:power}
\end{center}
\end{figure}

Summary statistics such as the mean intensity and the power spectrum
are primary tools to study the distribution and the properties of the emission-line galaxies. 
These can then be used 
for galaxy population studies or for
cosmological parameter inference.
In this section, we examine how well the mean intensity and the power spectrum are reconstructed.
We take medians of the reconstructed statistics by 5 different networks and compare them with true ones.

Fig. \ref{fig:mean_intensity} shows the correspondence of the true
and the reconstructed mean intensities. 
The mean intensities are widely distributed because of the cosmological 
variance of the underlying density field.
We see clear correlations between the true and the reconstructed mean intensities.
Fig. \ref{fig:mean_intensity} shows that the mean \Ha ([\OIII]) intensity of each $(1.7~\rm deg)^2$ map can be estimated with $\sim 10$\% ($\sim 20$\%) accuracy. 
We note that the residual uncertainty is comparable to those resulting from 
the luminosity function estimates by recent galaxy surveys \citep{Sobral13, Khostovan15}.
Planned LIM observations would have a much larger observational area.
For instance, SPHEREx \citep{Dore16} will perform
a deep survey over $200~\rm deg^2$
and thus the estimated mean intensities
would have a much smaller statistical uncertainty.

We test if our networks generate accurate images (intensity maps) if the
input observed map is significantly different from the training data. To this end, we input
intensity maps with the mean differing as much as 20 \%.
Some of these samples have mean intensities below or above the range plotted in Fig. 2.
We find that the networks reconstruct the \Ha and [\OIII] intensities with accuracy similar to those shown in Fig. \ref{fig:mean_intensity} when both maps are scaled with the same factor.
However, we find systematic offset when only one map is scaled more than 10\% while the other being unchanged. 
In order to reconstruct these ”outliers” accurately,
we need to consider a wide variety of training data, and/or to combine multiple
networks trained with maps in different mean intensity ranges.

Another important statistic is two-dimensional power spectrum.
Fig. \ref{fig:power} shows the variation of the true (shaded regions) and the reconstructed (error bars) power spectra.
For reference, we also show the power spectra of unsmoothed true maps.\footnote{
Power spectrum of the unsmoothed map $P(k)$ can be recovered from that of the smoothed map $P_{\rm sm}(k)$ by
$P(k) = \exp(k^2\sigma^2)P_{\rm sm}(k)$,
where $\sigma$ is the smoothing scale of the Gaussian beam.
}
Clearly, our networks {\it learn} the clustering of galaxies even though we do not explicitly teach that galaxies at different redshifts have different clustering amplitudes.
The top panel of Fig. \ref{fig:power} shows the difference between the true and the reconstructed power spectra normalized by the square root of the variance of the true power spectra $\sigma_{\rm true}$.
We note that the variance of the training data is also $\sigma_{\rm true}$.
For \Ha map, the difference is typically less than $\sigma_{\rm true}$ at large scales;
our network is able to recover the power spectrum of \Ha at $z=1.3$ with an accuracy of $\sim 10\%$ from a confused map.

\section{Discussion}

We have shown, for the first time, that cGANs can separate desired signals confused in an intensity map.
We can also locate intensity peaks where emission line galaxies are clustered at the target redshift.
Combining the distribution of the peaks and other information from follow-up observations of individual galaxies
would allow us to study the environmental dependence of the galaxy formation. 

A promising approach is to combine our deep learning method
with other conventional method such as cross-correlation analysis.
From the statistical information such as the power spectrum and the mean intensity of the reconstructed intensity maps (galaxy distributions) at a wide range of redshift, 
we can infer cosmological parameters and can also learn about the evolution of galaxy populations. 

In this letter, we have presented the results from our first attempt,
and there is much room for improvement.
In order for our method to be applied to real LIM observations, 
the networks need to be trained with
observational noises and other contaminants.
For cosmology studies, it would be important to train the networks 
with a variety of astrophysical/cosmological models and parameters 
to improve robustness.
Quantifying the uncertainty or the reliability of a reconstructed map is also important. 
When an input map is quite different from the training dataset, the networks should ideally return such information
together with the reconstructed map(s).
Methods shown in \citet{Kendall17} can be used for these purposes. 
To improve the ability of the networks,
we can utilize the three-dimensional 
information or train the networks with a larger survey area/volume.
Our networks can also be applied to observations in different wavelength 
such as sub-millimeter LIM.
We choose two emission lines with relatively close redshifts having similar structures in this study.
If one focus on two redshifts with larger separation, it could be easier for the networks to learn the difference and reproduce the maps well.
We continue exploring the deep learning approach to, for instance, de-noise intensity maps or to extract designated information from a map with more than two components.

\section*{acknowledgements}
We thank the anonymous referee for providing us useful comments.
We thank Yasuhiro Imoto for useful discussion. 
KM is supported by JSPS KAKENHI Grant Number 19J21379.
NF's visit at the University of Tokyo was supported by the
Princeton-UTokyo strategic partnership grant.
MS is supported by JSPS Overseas Research Fellowships. NY acknowledges financial support from JST CREST (JPMHCR1414).
A part of our computations in this study is carried out on Cray XC50 at Center for
Computational Astrophysics, National Astronomical Observatory of Japan.

\bibliography{bibtex_library} 

\label{lastpage}

\end{document}